\documentclass[10pt]{article}
\usepackage[T1]{fontenc}

\makeatletter

\newcommand{\LyX}{L\kern-.1667em\lower.25em\hbox{Y}\kern-.125emX\@}

\makeatother

\begin{document}

{\par\centering {\large Ahoronov Anandan Phase for the Quasi Exactly Solvable
Bose Systems}\large \par}
\vspace{0.6cm}

{\par\centering \textbf{\small ANIRBAN PATHAK}\small \par}
\vspace{0.2cm}

{\par\centering \textit{\small Department of Physics}\small \par}

{\par\centering \textit{\small Visva Bharati }\small \par}

{\par\centering \textit{\small Santiniketan-731235}\small \par}
\vspace{0cm}

{\par\centering \textit{\small India}\small \par}
\vspace{1cm}

\begin{abstract}
An extended notion of quasi-exactly solvable potential model is used here to
treat quasi exactly solvable (QES) Bose systems. We report an analytic expression
for the Ahoronov Anandan non-adiabatic geometric phase for the QES Bose system
in general. The generalized expression is then used to study some particular
cases of physical interest and we observe that the geometric phase can be tuned. 
\end{abstract}

\section{Introduction:}

Phase is responsible for all the interference phenomenon we observe in classical
and quantum physics. So people have studied the properties of the quantum phase
from different contexts since the beginning of quantum mechanics{[}\ref{perinova}{]}.
But the interest increased considerably in the recent past when Berry published
his famous paper {[}\ref{Berry}{]}. He showed that the phase acquired by the
quantum system during a cyclic evolution under the action of an adiabatically
varying Hamiltonian is a sum of two parts. The first part is dynamical in nature
and the second is geometric. Later on Ahoronov and Anandan generalized the idea
of the Berry phase and they defined a geometric phase factor for any cyclic
evolution of quantum system {[}\ref{Aharonov}{]}. The existence of geometric
phase is found in many areas of physics. The examples are from the quantum Hall
effect to the Jahn-Teller effect and from the spin orbit interaction to quantum
computation {[}\ref{perinova},\ref{Joshi-Pati},\ref{blais}{]}. 

Physicists have always tried to understand Nature in terms of simple models
and the simple harmonic oscillator (SHO) is probably the most important one
among those models. The SHO model can be used to understand a wide variety of
physical phenomena ranging from problems in Newtonian mechanics to those in
quantum field theory. However, for a real physical problem one has to incorporate
anharmonicity in the model Hamiltonian. For example, it is related to quantum
Bose models with interaction or self-interaction. The Schrodinger equation for
these anharmonic Bose oscillators are not exactly solvable. However, Dolya and
Zaslavaskii {[}\ref{Dolya}{]} have recently shown that the notion of the quasi
exactly solvable (QES) models can be extended for anharmonic Bose oscillators.
They have reported that a class of Hamiltonian, often faced by us in molecular
physics, theory of magnetism and other brunches of physics is QES. This fact
provoked us to study the possibilities of observing non-adiabatic geometric
phase or Ahoronov-Anandan phase for the quasi exactly solvable Bose systems.
In the present paper we have found an analytic expression in closed form for
the Ahoronov-Anandan phase for the QES bose systems characterized by the Hamiltonian
defined by Dolya and Zaslavaskii {[}\ref{Dolya}{]}. The generalized expression
is then used to study a particular case of physical interest. 

In section 2 we give a short introduction to the Ahoronov-Anandan phase{[}\ref{Aharonov},\ref{Moore}{]}.
We obtain a generalized expression for geometric phase of QES Bose systems in
section 3 and use that expression to study some particular cases of physical
interest in the subsequent subsection. We devote section 4 for discussions and
conclusions.

\section{Ahoronov-Anandan Phase}

In a quantum mechanical description of any system, one has a complex Hilbert
space $\cal{H}$ whose nonzero vectors represent states of the system. A vector
\( |\psi > \) in $\cal{H}$ and any complex multiple of it (i.e \( c|\psi > \))
represent the same state in the Hilbert space $\cal{H}$. This arbitrariness
in the representation of states can be reduced by imposing the normalization
condition 
\begin{equation}
\label{ek}
<\psi |\psi >=1.
\end{equation}
 There still remains an arbitrariness of a phase factor because a normalized
vector \( |\psi > \) and the vectors \( \alpha |\psi > \) represent the same
state in Hilbert space provided that the modulus of \( \alpha  \) is unity
(i.e \( |\alpha |=1 \)). The collection of all vectors \( \alpha  \)\( |\psi > \)
with \( |\psi > \), a fixed normalized vector and \( \alpha  \) taking all
possible complex values of modulus one is called a unit ray and is denoted by
\( |\widetilde{\psi }> \){[}\ref{Tulshidas}{]}. More general rays consist
of a collection of all vectors of the form \( c|\psi > \). So rays represent
vectors without any arbitrariness of phase. 

To understand the geometric phase let us start from a simple situation in which
a state vector \( |\psi > \) evolves cyclically in the Hilbert space $\cal{H}$
and after a complete period \( T \) returns to the same state vector with a
different phase which may be written as \( \alpha |\psi >=\exp (i\phi )|\psi > \).
Therefore, 
\begin{equation}
\label{tin}
|\psi (T)>=\exp \left( i\phi \right) |\psi (0)>,
\end{equation}
 where \( \phi  \) is the total phase which has two parts. The first part is
the dynamical phase and the second is the geometric phase. We are interested
in the later type of phase which can be found out by subtracting the dynamical
phase part from the total phase. 

The cyclic evolution of the state vector may be described by a curve \( C \)
in $\cal{H}$ because \( |\psi > \) and \( \alpha |\psi > \) are two different
points in $\cal{H}$. Since there are infinitely many possible values of \( \alpha  \)
so there are infinitely many curves \( C \) in $\cal{H}$ which describes the
same cyclic evolution. In the projective Hilbert space of rays \( P \) or in
the ray-representative space all these points are represented by the same ray
(i.e by the same point). So we will have a single closed curve \( \widehat{C} \)
in the ray representation space \( P \) corresponding to the infinitely many
possible curves in the usual Hilbert space. As there is no arbitrariness of
phase in \( P \) so in the interval \( [0,T] \) we must have 
\begin{equation}
\label{four}
|\widetilde{\psi }(T)>=|\widetilde{\psi }(0)>.
\end{equation}
 Ahoronov and Anandan{[}\ref{Aharonov}{]} exploited this property of single
valuedness of the unit ray in the ray representative space \( P \) to establish
the presence of a geometric phase for all cyclic evolutions. They started by
supposing that the normalized state \( |\psi (t)>\in  \)$\cal{H}$ evolves according
to the Schr\( \ddot{o} \)dinger equation 
\begin{equation}
\label{dui}
H(t)|\psi (t)>=i\hbar \left( \frac{d}{dt}|\psi (t)>\right) .
\end{equation}
 We can always define a state vector \( |\widetilde{\psi }(t)> \) in \( P \)
which is equivalent to \( |\psi (t)> \) as 
\begin{equation}
\label{char}
|\widetilde{\psi }(t)>=\exp (-if(t))|\psi (t)>.
\end{equation}
 Now substituting (\ref{tin} and \ref{char}) in (\ref{four}) we obtain 
\begin{equation}
\label{5}
f(T)-f(0)=\phi .
\end{equation}

If we assume that the total wave function rotates by \( 2\pi  \) radian for
this \( T \) which results in a cyclic motion of every state vector of the
Hilbert space $\cal{H}$, then we will have another condition
\begin{equation}
\label{cond1}
f(T)-f(0)=2\pi .
\end{equation}
 Mere satisfaction of condition (\ref{cond1}) will ensure the single valuedness
of the wave function in the usual Hilbert space $\cal{H}$\emph{.} Again from
(\ref{dui} and \ref{char}) we have 
\begin{equation}
\label{choi.1}
H|\psi (t)>=i\hbar \left( \frac{d}{dt}\exp (if(t))|\widetilde{\psi }(t)>\right) 
\end{equation}
 or, 
\begin{equation}
\label{choi.2}
H|\psi (t)>=-\hbar \frac{df}{dt}\exp (if(t))|\widetilde{\psi }(t)>+i\hbar \exp (if(t))\frac{d}{dt}|\widetilde{\psi }(t)>
\end{equation}
 Since \( \psi (t) \) is normalized so we have 
\begin{equation}
\label{choi}
-\frac{df}{dt}=\frac{1}{\hbar }<\psi (t)|H|\psi (t)>-<\widetilde{\psi }(t)|i\frac{d}{dt}|\widetilde{\psi }(t)>.
\end{equation}
 Now integrating equation (\ref{choi}) from \( 0 \) to \( T \) and using
(\ref{5}) we have 
\begin{equation}
\label{7}
\phi =-\frac{1}{\hbar }\int ^{T}_{0}<\psi (t)|H|\psi (t)>dt+\int ^{T}_{0}<\widetilde{\psi }(t)|i\frac{d}{dt}|\widetilde{\psi }(t)>dt=\gamma +\beta 
\end{equation}
 where 
\begin{equation}
\label{8}
\gamma =-\frac{1}{\hbar }\int ^{T}_{0}<\psi (t)|H|\psi (t)>dt
\end{equation}
 is proportional to the time integral of the expectation value of the Hamiltonian.
It is dynamical in origin and hence called dynamical phase. Now if we subtract
the dynamical phase part \( \gamma  \) from the total phase \( \phi  \) then
we will get the geometric phase or the Ahoronov Anandan phase
\begin{equation}
\label{9}
\beta =\int ^{T}_{0}<\widetilde{\psi }(t)|i\frac{d}{dt}|\widetilde{\psi }(t)>dt.
\end{equation}
 It is clear that the same \( |\widetilde{\psi }(t)> \) can be chosen for infinitely
many possible curves \( C \) by appropriate choice of \( f(t) \). This phase
is geometrical in the sense that it does not depend on the Hamiltonian responsible
for the evolution. Moreover it is independent of the parameterization and redefinition
of phase of \( |\psi (t)> \). Hence, \( \beta  \) depends only on the evolution
of the shadow of \( |\psi (t)> \) in projective Hilbert space \( P \) and
is a pure geometrical entity {[}\ref{Aharonov}-\ref{Moore}{]}.

\section{Quasi exactly solvable anharmonic oscillator:}

In the introduction we have stated that the QES anharmonic Bose oscillators
are very common in Nature. Therefore, a generalized treatment for the geometric
phase of the QES anharmonic Bose oscillator may find its application in different
directions of physics. Keeping these facts in mind, we will study the possibilities
of observing Ahoronov-Anandan phase for an QES anharmonic Bose oscillator potential
in general.

The Hamiltonian of a generalized QES anharmonic Bose oscillator defined by Dolya
and Zaslavaskii {[}\ref{Dolya}{]} is 
\begin{equation}
\label{ten}
\begin{array}{lcl}
H & = & H_{0}+V\\
 & = & \sum ^{p_{0}}_{p=1}\epsilon _{p}(a^{\dagger }a)^{p}+\sum ^{s_{0}}_{s=0}A_{s}\left[ (a^{\dagger }a)^{s}a^{2}+a^{\dagger ^{2}}(a^{\dagger }a)^{s}\right] 
\end{array}
\end{equation}
 where \( a \) and \( a^{\dagger } \) are the usual annihilation and creation
operators respectively. Here we have chosen to work in units in which \( m=1,\, \omega =1,\, \hbar =1. \)
Details of the quasi exactly solvability criterion will be available in the
work of Dolya and Zaslavaskii {[}\ref{Dolya}{]}. Now the total eigen state
of the unperturbed Hamiltonian \( H_{0} \) can be written as 
\begin{equation}
\label{new1}
|\psi (0)>=\sum _{n}C_{n}|n>
\end{equation}
 where the coefficients \( C_{n} \)'s depend on the initial conditions. The
total wave function of the Hamiltonian \( H \) satisfying the Schr\( \ddot{o} \)dinger
equation (\ref{dui}) is 
\begin{equation}
\label{label2}
\begin{array}{lcl}
|\psi (t)> & = & \sum _{n}C_{n}\exp \left( -iHt\right) |n>\\
 & = & \sum _{n}C_{n}\exp \left( \sum ^{p_{0}}_{p=1}-i\epsilon _{p}(a^{\dagger }a)^{p}t+\sum ^{s_{0}}_{s=0}-iA_{s}\left[ (a^{\dagger }a)^{s}a^{2}+a^{\dagger ^{2}}(a^{\dagger }a)^{s}\right] t\right) |n>.
\end{array}
\end{equation}
 For the calculation of the geometric phase, we have to chose the single-valued
state \( |\widetilde{\psi }(t)> \). This can be easily done by choosing a \( f(t) \)
which simultaneously satisfies equation (\ref{four} and \ref{cond1}). The
simplest choice is \( f(t)=\lambda t \), where \( \lambda =\sum ^{s_{0}}_{s=0}A_{s} \)
. Now with that choice we have 
\begin{equation}
\label{new7}
\begin{array}{lcl}
|\widetilde{\psi }(t)> & = & \exp (-if(t))|\psi (t)>\\
 & = & \sum _{n}C_{n}\exp \left( \sum ^{p_{0}}_{p=1}-i\epsilon _{p}(a^{\dagger }a)^{p}t+\sum ^{s_{0}}_{s=0}-iA_{s}\left( 1+\left[ (a^{\dagger }a)^{s}a^{2}+a^{\dagger ^{2}}(a^{\dagger }a)^{s}\right] \right) t\right) |n>
\end{array}
\end{equation}
 and 
\begin{equation}
\label{new7.1}
f(T)=\lambda T=2\pi 
\end{equation}
Therefore, from equations (\ref{9} ,\ref{new7} and \ref{new7.1}) a compact
expression for the Ahoronov Anandan phase of a QES Bose anharmonic system can
be obtained as 
\begin{equation}
\label{new8}
\begin{array}{lcl}
\beta  & = & \int ^{T}_{0}<\widetilde{\psi }(t)|i\frac{d}{dt}|\widetilde{\psi }(t)>dt\\
 & = & \int ^{T}_{0}\sum _{k}\sum _{n}<k|C_{k}^{*}C_{n}\exp \left( \sum ^{p_{0}}_{p=1}i\epsilon _{p}(a^{\dagger }a)^{p}t+\sum ^{s_{0}}_{s=0}iA_{s}\left( 1+\left[ (a^{\dagger }a)^{s}a^{2}+a^{\dagger ^{2}}(a^{\dagger }a)^{s}\right] \right) t\right) \\
 & \times  & \left( \sum ^{p_{0}}_{p=1}\epsilon _{p}(a^{\dagger }a)^{p}+\sum ^{s_{0}}_{s=0}A_{s}\left( 1+\left[ (a^{\dagger }a)^{s}a^{2}+a^{\dagger ^{2}}(a^{\dagger }a)^{s}\right] \right) t\right) \\
 & \times  & \exp \left( \sum ^{p_{0}}_{p=1}-i\epsilon _{p}(a^{\dagger }a)^{p}t+\sum ^{s_{0}}_{s=0}-iA_{s}\left( 1+\left[ (a^{\dagger }a)^{s}a^{2}+a^{\dagger ^{2}}(a^{\dagger }a)^{s}\right] \right) t\right) |n>dt\\
 & = & \int ^{T}_{0}\sum _{k}\sum _{n}<k|C_{k}^{*}C_{n}\left( \sum ^{p_{0}}_{p=1}\epsilon _{p}(a^{\dagger }a)^{p}+\sum ^{s_{0}}_{s=0}A_{s}\left( 1+\left[ (a^{\dagger }a)^{s}a^{2}+a^{\dagger ^{2}}(a^{\dagger }a)^{s}\right] \right) \right) |n>dt\\
 & = & 2\pi +T\sum _{n}|C_{n}|^{2}\sum ^{p_{0}}_{p=1}\epsilon _{p}(n)^{p}+T\sum _{n}C_{n-2}^{*}C_{n}\sum ^{s_{0}}_{s=0}A_{s}(n-2)^{s}\sqrt{n(n-1)}\\
 & + & T\sum _{n}C_{n+2}^{*}C_{n}\sum ^{s_{0}}_{s=0}A_{s}(n)^{s}\sqrt{(n+1)(n+2)}.\\
 &  & \\
 &  & \\
 &  & \\
 &  & \\
 &  & 
\end{array}
\end{equation}

The quantum nature of the geometric phase is manifested here through the discrete
sum appearing in the expression (\ref{new8}) and the state dependence of the
phase appears through the coefficients \( C_{k}^{*}C_{n} \). Now let us think
of some special situations of physical interest.

\subsection{Input is an intense laser beam.}

Let us now consider the case where an intense electromagnetic field having Poissonian
statistics (\( C_{n}=\exp (-\frac{|\alpha |^{2}}{2})\frac{\alpha ^{n}}{\sqrt{n!}} \))
interacts with a system characterized by the Hamiltonian (\ref{ten}). With
the help of equations (\ref{new8}) the geometric phase \( \beta  \) for this
physical system can be written as 
\[
\begin{array}{lcl}
\beta  & = & 2\pi +T\sum _{n}|C_{n}|^{2}\sum ^{p_{0}}_{p=1}\epsilon _{p}(n)^{p}+T\sum _{n}C_{n-2}^{*}C_{n}\sum ^{s_{0}}_{s=0}A_{s}(n-2)^{s}\sqrt{n(n-1)}\\
 & + & T\sum _{n}C_{n+2}^{*}C_{n}\sum ^{s_{0}}_{s=0}A_{s}(n)^{s}\sqrt{(n+1)(n+2)}\\
 & = & 2\pi +T\exp \left( -|\alpha |^{2}\right) \sum _{n}\frac{|\alpha |^{2n}}{n!}\left( \sum ^{p_{0}}_{p=1}\epsilon _{p}(n)^{p}+\sum ^{s_{0}}_{s=0}2|\alpha |^{2}A_{s}(n)^{s}\cos (2\theta )\right) .\\
 & 
\end{array}\]
where \( \alpha =|\alpha |\exp (i\theta ) \). Here we observe that the geometric
phase depends on the phase of the initial laser beam. The presence of the off-diagonal
terms in the interaction Hamiltonian is manifested through the presence of \( \theta  \)
dependent terms in the expression for the geometric phase \( \beta  \).

\section{Discussions and Conclusions:}

\begin{table}
{\centering \begin{tabular}{|l|c|}
\hline 
{\small Photon Statistics}&
{\small \( \, \, \, \, \, \, \, \, \, |C_{n}|^{2} \)}\\
\hline 
{\small Sub-Poissonian}&
{\small \( \frac{N!}{(N-n)!n!}p^{n}(1-p)^{N-n}\, \, (0\leq p\leq 1) \)}\\
\hline 
{\small Poissonian}&
\( \exp (-|\alpha |^{2})\frac{\alpha ^{n}}{\sqrt{n!}} \)\\
\hline 
{\small Super-Poissonian}&
{\small \( \frac{(n+W-1)!}{(W-1)!n!}q^{n}(1-q)^{W}\, \, (0\leq q\leq 1) \)}\\
\hline 
\end{tabular}\small \par}
{\par\centering table1\par}
\end{table}

In the present work we have seen that the geometric phase appears in QES Bose
systems and depends on the photon statistics of the input field through \( C_{n} \)'s
(see table 1). We have also observed that the presence of the off-diagonal terms
in the interaction Hamiltonian is manifested through the presence of \( \theta  \)
dependent terms in the expression for the geometric phase \( \beta  \). From
these facts we can conclude two things, firstly, one should always have to take
care of the possibilities of appearance of geometric phase while working with
QES Bose systems and, secondly, geometric phase can be tuned because it depends
strongly on the phase of the input field which can be tuned. 

~\\
\textbf{Acknowledgement:} Author is thankful to CSIR, India for a senior research
fellowship. 
\vspace{2cm}

\textbf{References:}

\begin{enumerate}
\item {\footnotesize \label{perinova}Perinova V, Luks A and Perina J 1998} \emph{\footnotesize Phases
in Optics} {\footnotesize (New Jersey: World Scientific ) }{\footnotesize \par}
\item {\footnotesize \label{Berry}Berry M V 1984} \emph{\footnotesize Proc. R. Soc
A} \textbf{\footnotesize 392} {\footnotesize 45.}{\footnotesize \par}
\item {\footnotesize \label{Aharonov}Ahoronov Y and Anandan J 1987} \emph{\footnotesize Phys.
Rev. Lett.}{\footnotesize ,} \textbf{\footnotesize 58} {\footnotesize 1593.}{\footnotesize \par}
\item {\footnotesize \label{Joshi-Pati}Joshi A, Pati A and Banerjee A 1994} \emph{\footnotesize Phys.
Rev. A} {\footnotesize 49 5131.}{\footnotesize \par}
\item {\footnotesize \label{blais}Blais A and Tremblay A M S} \emph{\footnotesize Preprint}
{\footnotesize quant-ph/0105006.}{\footnotesize \par}
\item {\footnotesize \label{Dolya}Dolya S N and Zaslavskii O B 2000} \emph{\footnotesize J.
Phys. A} \textbf{\footnotesize 33} {\footnotesize L369.}{\footnotesize \par}
\item {\footnotesize \label{Moore}Moore D J 1991} \emph{\footnotesize Phys. Rep.}
\textbf{\footnotesize 210} {\footnotesize 1.}{\footnotesize \par}
\item {\footnotesize \label{Tulshidas}Dass T and Sharma S K 1998} \emph{\footnotesize Mathematcal
Methods in Classical and Quantum Physics} {\footnotesize (Hyderabad: Uniiversity
Press) }{\footnotesize \par}
\end{enumerate}
\end{document}